%% file: ArXivVersion.tex
\documentclass[11pt,preprint]{aastex}

\usepackage{gensymb}
\usepackage{amsmath}
\usepackage{bm}
\usepackage{natbib}

\input{mycommands}
\shortauthors{Philip Engelke}

\begin{document}
\bibliographystyle{plainnat}


\title{
MOND Fit of Nature Physics 11:245 Mass Distribution Model to Rotation Curve Data}




\author{Philip D. Engelke}
\author{Department of Physics and Astronomy, The Johns Hopkins University}

%
%



In a recent note, \citet{ioc15b} analyze the consistency of Modified Newtonian Dynamics (MOND) with their compiled Milky Way data and baryonic mass distribution models from \citet{ioc15a}, looking especially at whether they recover the canonical value of the MOND critical acceleration $a_0$ parameter when fitting two alternate versions of the MOND function using the value of the MOND constant $a_0$ as an adjustable parameter.  In this way, they tested the "standard" interpolation function (the original, proposed by \citet{mil83}) and the "simple" interpolation function \citep{fam05}.  What they report finding is that the standard interpolation function requires a different value of $a_0$ from that used for external galaxies in order to fit their Milky Way data, whereas the simple interpolation function can fit the observed rotation curve for "a subset of models" \citep{ioc15b} using the traditional $a_0$ value.

However, they do not explicitly show in their paper a plot of the resultant MOND fit through the rotation curve data.  We plot this  using the simple interpolation function, which has been found in a comparison against the standard interpolation to give better fits to the rotation curves of a number of external galaxies \citep{gen11}.  We read 450 points from the Newtonian rotation curve shown as a thin black line directly off of Figure 2 in \citet{ioc15a} using online image pixel coordinate software, and fed the values into the MOND formula.  The result, when superimposed on the same figure, strikingly passes right through the red points showing the compilation of real observational Milky Way rotation curve data.  We display these results in Figure 1 of this paper, with our MOND rotation curve prediction plotted in green on top of a copy of Figure 2 from \citet{ioc15a}.           


\begin{figure*}[ht!] 
\vspace{0.1in}
\begin{center}
\includegraphics[width=1.0\textwidth, angle=0]{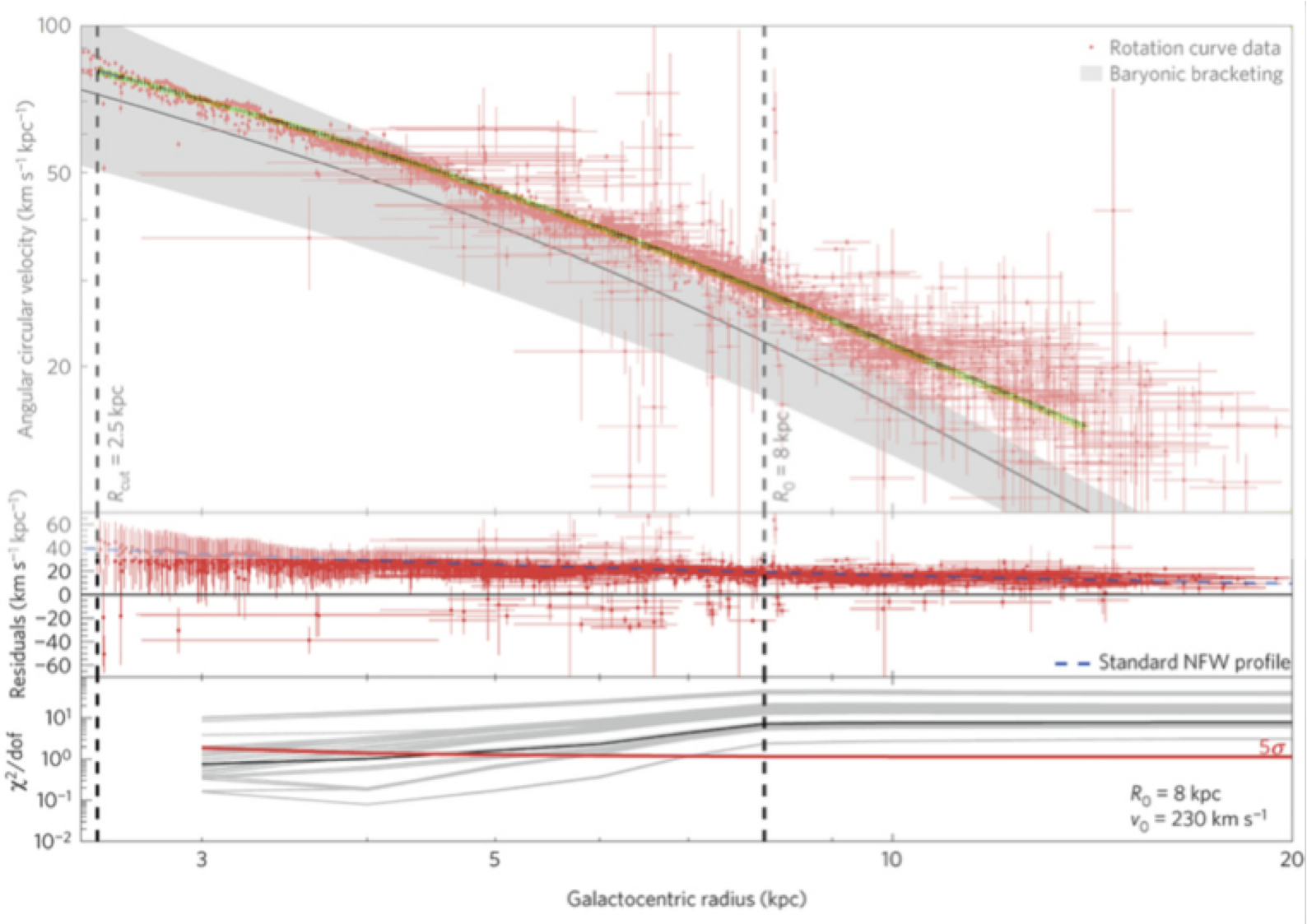} \hspace{0.2in} 
\caption{Our predicted MOND rotation curve plotted in green on top of a copy of Figure 2 from \citet{ioc15a}.  In their figure, the dark narrow curve shows the predicted Newtonian rotation curve prediction using one of the baryonic mass distributions considered by \citet{ioc15a}, and the gray region represents the margin of uncertainty in the Newtonian rotation curve prediction due to uncertainties from all seventy of the baryonic models that they considered.  The red data points are their measured rotation curve data, from the compilation of sources. \citep{ioc15a}.  Note that our green MOND fit using the simple interpolation function and their example baryonic mass distribution model fits their compiled data remarkably well despite no parameter adjustments.}
\end{center}
\end{figure*}


Our MOND rotation curve prediction was calculated as follows.  We begin with the MOND equation for a general interpolation function $\mu(x)$, where $x = a/a_0$ and $a_0 = 1.2 \times 10^{-10}$ m/s$^2$.  
\begin{equation}
a = \frac{F}{m\mu(x)}.  
\end{equation}
Defining $g$ as the acceleration predicted under Newtonian mechanics and gravity, the acceleration $a$ predicted by MOND is given by
\begin{equation}
a\mu(x) = g.
\end{equation}
The simple interpolation function is given by
\begin{equation}
\mu(x) = \frac{x}{1 + x}
\end{equation}
which, when inserted into the MOND acceleration equation, yields
\begin{equation}
a\frac{\frac{a}{a_0}}{1 + \frac{a}{a_0}} = g.
\end{equation}
Rearranging the equation, we are faced with a quadratic equation in a:
\begin{equation}
a^2 - ga - ga_0 = 0.
\end{equation}
Solving for a, we find
\begin{equation}
a = \frac{g + \sqrt{g^2 + 4ga_0}}{2}.
\end{equation}
To find the circular velocities, we write
\begin{equation}
a = \frac{v_c^2}{R}
\end{equation}
which means that
\begin{equation}
v_c = \sqrt{aR}.
\end{equation}

\citet{ioc15a} plot angular circular velocity $v_c / R$ in terms of km/s per kpc, so to compare the MOND predicted circular velocities to the data that they plotted, we find the MOND angular circular velocities in the same units as they do by converting the circular velocities into km/s and dividing by the distance from the Galactic center in kpc.  

This one visually striking fit is of course not a conclusive demonstration that MOND using the simple interpolation function describes the rotation curve of the Milky Way, because the baryonic mass distribution shown as a thin black line in \citet{ioc15a} Figure 2 is but one of many possible models, and there could be systematic errors in the compilation of the data, as pointed out by \citet{mcg15}.  The analysis performed in \citet{ioc15b} is valuable and extensive, and shows that MOND with the simple interpolation function cannot be ruled out as a fit to the Milky Way rotation curve.  However, we believe that this visual fit to the rotation curve data is insightful as a supplement to the reports of \citet{ioc15b}.  The results are consistent with the findings of \citet{mcg08}, a similar MOND fit to the Milky Way rotation curve.  They are also consistent with studies of MOND fits to other galaxies using the simple interpolation function, such as \citet{gen11}.  

NOTE: Several corrections have been made in this update to arXiv:1505.06174.  We apologize for the previous mischaracterizations.

\clearpage

\bibliography{mybib}

\end{document}

%% file: mycommands.tex
\newcommand{\lsim}{\mbox{$\mathrel{\vcenter{\hbox{\ooalign{\raise3pt\hbox{$<$}\crcr \lower3pt\hbox{$\sim$}}}}}$}}
\newcommand{\gsim}{\mbox{$\mathrel{\vcenter{\hbox{\ooalign{\raise3pt\hbox{$>$}\crcr \lower3pt\hbox{$\sim$}}}}}$}}